\documentclass[]{spie}  

\usepackage{amsmath,amsfonts,amssymb}
\usepackage{textcomp}
\usepackage{graphicx}
\usepackage[colorlinks=true, allcolors=blue]{hyperref}

\title{Overview and status of EXCLAIM, the experiment for cryogenic large-aperture intensity mapping}

\author[a]{Giuseppe Cataldo}
\author[b]{Peter Ade}
\author[a]{Christopher Anderson}
\author[a]{Alyssa Barlis}
\author[a]{Emily Barrentine}
\author[a]{Nicholas Bellis}
\author[c]{Alberto Bolatto}
\author[d]{Patrick Breysse}
\author[a]{Berhanu Bulcha}
\author[e]{Jake Connors}
\author[a]{Paul Cursey}
\author[a]{Negar Ehsan}
\author[a]{Thomas Essinger-Hileman}
\author[a]{Jason Glenn}
\author[f]{Joseph Golec}
\author[a]{James Hays-Wehle}
\author[a]{Larry Hess}
\author[a]{Amir Jahromi}
\author[a]{Mark Kimball}
\author[a]{Alan Kogut}
\author[a]{Luke Lowe}
\author[g]{Philip Mauskopf}
\author[f]{Jeffrey McMahon}
\author[a]{Mona Mirzaei}
\author[h]{Harvey Moseley}
\author[c]{Jonas Mugge-Durum}
\author[a]{Omid Noroozian}
\author[i]{Trevor Oxholm}
\author[c]{Ue-Li Pen}
\author[j]{Anthony Pullen}
\author[a]{Samelys Rodriguez}
\author[a]{Peter Shirron}
\author[i]{Gage Siebert}
\author[g]{Adrian Sinclair}
\author[k]{Rachel Somerville}
\author[g]{Ryan Stephenson}
\author[a]{Thomas Stevenson}
\author[a]{Eric Switzer}
\author[i]{Peter Timbie}
\author[b]{Carole Tucker}
\author[l]{Eli Visbal}
\author[d]{Carolyn Volpert}
\author[a]{Edward Wollack}
\author[j]{Shengqi Yang}

\affil[a]{NASA Goddard Space Flight Center, Greenbelt, MD, USA}
\affil[b]{Cardiff University, Cardiff, UK}
\affil[c]{University of Maryland, College Park, MD, USA}
\affil[d]{Canadian Institute for Theoretical Astrophysics, Toronto, Canada}
\affil[e]{National Institute of Standards and Technology, Boulder, CO, USA}
\affil[f]{University of Chicago, Chicago, IL, USA}
\affil[g]{Arizona State University, Phoenix, AZ, USA}
\affil[h]{Quantum Circuits, New Haven, CT, USA}
\affil[i]{University of Wisconsin-Madison, Madison, WI, USA}
\affil[j]{New York University, New York, NY, USA}
\affil[k]{Rutgers University, New Brunswick, NJ, USA}
\affil[l]{University of Toledo, Toledo, OH, USA}

\authorinfo{Further author information: (Send correspondence to G.C.)\\G.C.: E-mail: giuseppe.cataldo@nasa.gov, Telephone: +1 301 286 3491}

\def\bossarea{320~${\rm deg}^2$}
\def\galaxyarea{${\sim}100~{\rm deg}^2$}
\def\exclaimz{$0 < z < 3.5$}
\def\exclaimband{$420$--$540$~GHz}
\def\exclaimr{$R=512$}

\def\mspec{\textmu-Spec}

\usepackage[normalem]{ulem}
\usepackage{color}

\begin{document} 
\maketitle

\begin{abstract}
The EXperiment for Cryogenic Large-Aperture Intensity Mapping (EXCLAIM) is a balloon-borne far-infrared telescope that will survey star formation history over cosmological time scales to improve our understanding of why the star formation rate declined at redshift $z < 2$, despite continued clustering of dark matter. Specifically, EXCLAIM will map the emission of redshifted carbon monoxide and singly-ionized carbon lines in windows over a redshift range $0 < z < 3.5$, following an innovative approach known as intensity mapping. Intensity mapping measures the statistics of brightness fluctuations of cumulative line emissions instead of detecting individual galaxies, thus enabling a blind, complete census of the emitting gas. To detect this emission unambiguously, EXCLAIM will cross-correlate with a spectroscopic galaxy catalog.
The EXCLAIM mission uses a cryogenic design to cool the telescope optics to approximately $1.7$~K. The telescope features a $90$-cm primary mirror to probe spatial scales on the sky from the linear regime up to shot noise-dominated scales. The telescope optical elements couple to six \textmu-Spec spectrometer modules, operating over a $420$--$540$\,GHz frequency band with a spectral resolution of $512$ and featuring microwave kinetic inductance detectors. A Radio Frequency System-on-Chip (RFSoC) reads out the detectors in the baseline design. The cryogenic telescope and the sensitive detectors allow EXCLAIM to reach high sensitivity in spectral windows of low emission in the upper atmosphere. Here, an overview of the mission design and development status since the start of the EXCLAIM project in early 2019 is presented. 
\end{abstract}


\section{Introduction}
Observations of our universe over cosmological time~\cite{2014ARA&A..52..415M, 2020ApJ...902..111W} indicate an increase in the rate of star formation from the period of cosmological reionization up to redshifts $z{\approx}2$. Following this peak, the star formation rate is inferred to fall by a factor of $10$, all while dark matter continues to cluster $100$-fold~\cite{2013ARA&A..51..105C}. To gain a better understanding of the aggregate evolution of galaxies in this cosmological context, new measurements are needed of the typical abundance, excitation and evolution of the molecular gas that forms and is influenced by stars. The EXperiment for Cryogenic Large-Aperture Intensity Mapping (EXCLAIM) is a high-altitude balloon-borne telescope that will address this scientific need by mapping the sub-millimeter emission of redshifted carbon monoxide (CO) and [CII] lines in windows varying over \exclaimz.

Rather than detecting individual galaxies, EXCLAIM will measure the statistics of brightness fluctuations of redshifted, cumulative line emission, an approach known as intensity mapping (IM)\cite{1979MNRAS.188..791H, 2019BAAS...51c.101K}. IM is sensitive to the integral of the luminosity function in cosmologically large volumes and to tracers of several environments in the interstellar medium (ISM). Current simulations~\cite{2019ApJ...882..137P} show tensions with measurements of CO in individual galaxies, thus motivating a blind, complete survey over a large area. 

In its baseline survey, EXCLAIM will map emission over a \exclaimband\ frequency band with resolving power \exclaimr\ on a \bossarea\ region of the sky that overlaps with the Baryon Oscillation Spectroscopic Survey (BOSS)~\cite{2020arXiv200708991E} and several \galaxyarea\ regions in the plane of the Milky Way. Even though the optimal band for BOSS cross-correlation is $420$--$600$~GHz, in EXCLAIM this is truncated at $540$~GHz to avoid bright ortho-water emission in the upper atmosphere at $557$~GHz. The broad survey area provides access to linear density fluctuations, which are easier to interpret than clustering within halos. EXCLAIM's primary extragalactic science uses cross-correlation to facilitate unambiguous detection of redshifted emission in the presence of foregrounds.

\begin{figure*}
\begin{center}
\begin{tabular}{ll}
\includegraphics[width=0.47\textwidth]{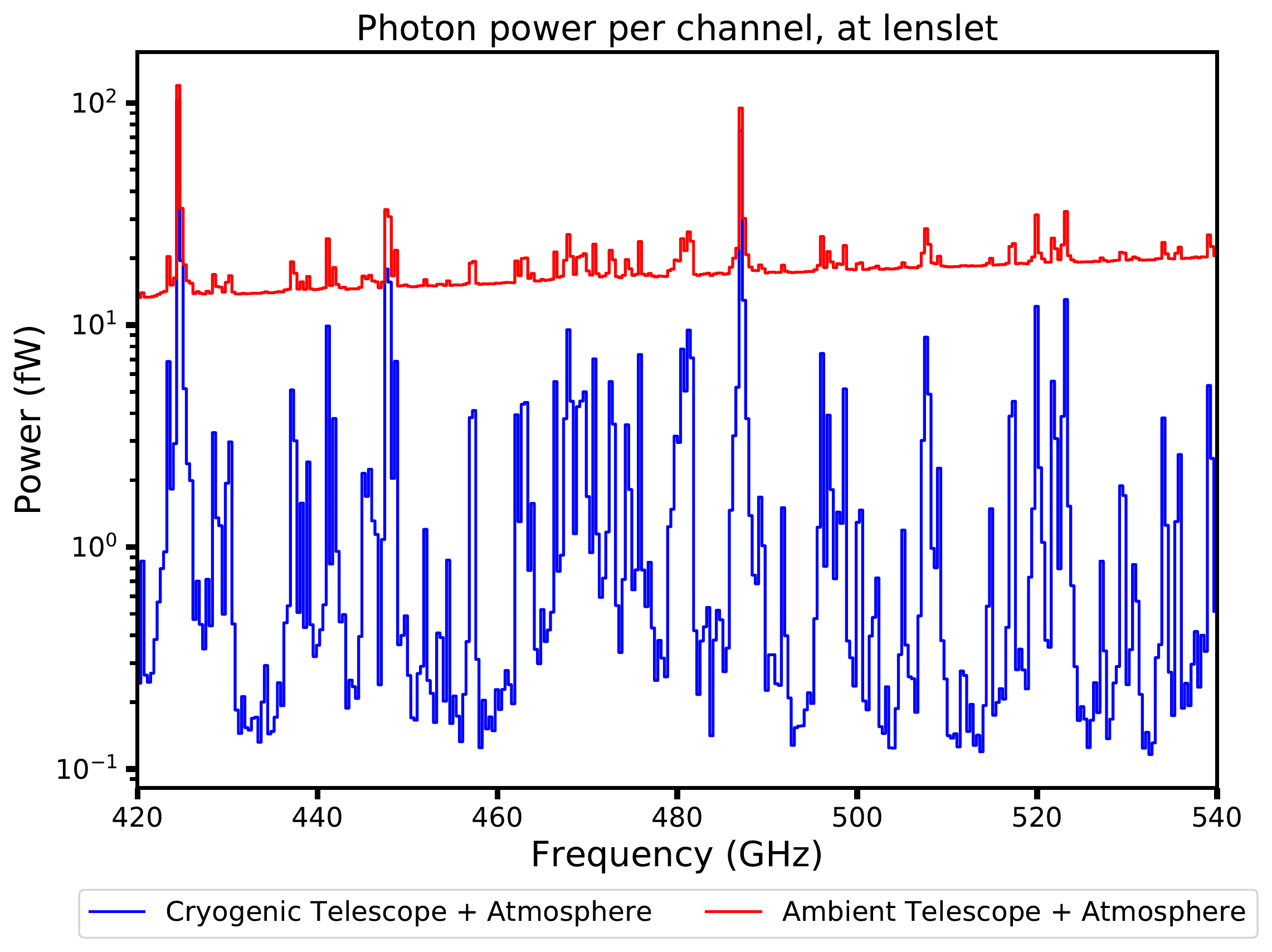} &
\includegraphics[width=0.48\textwidth]{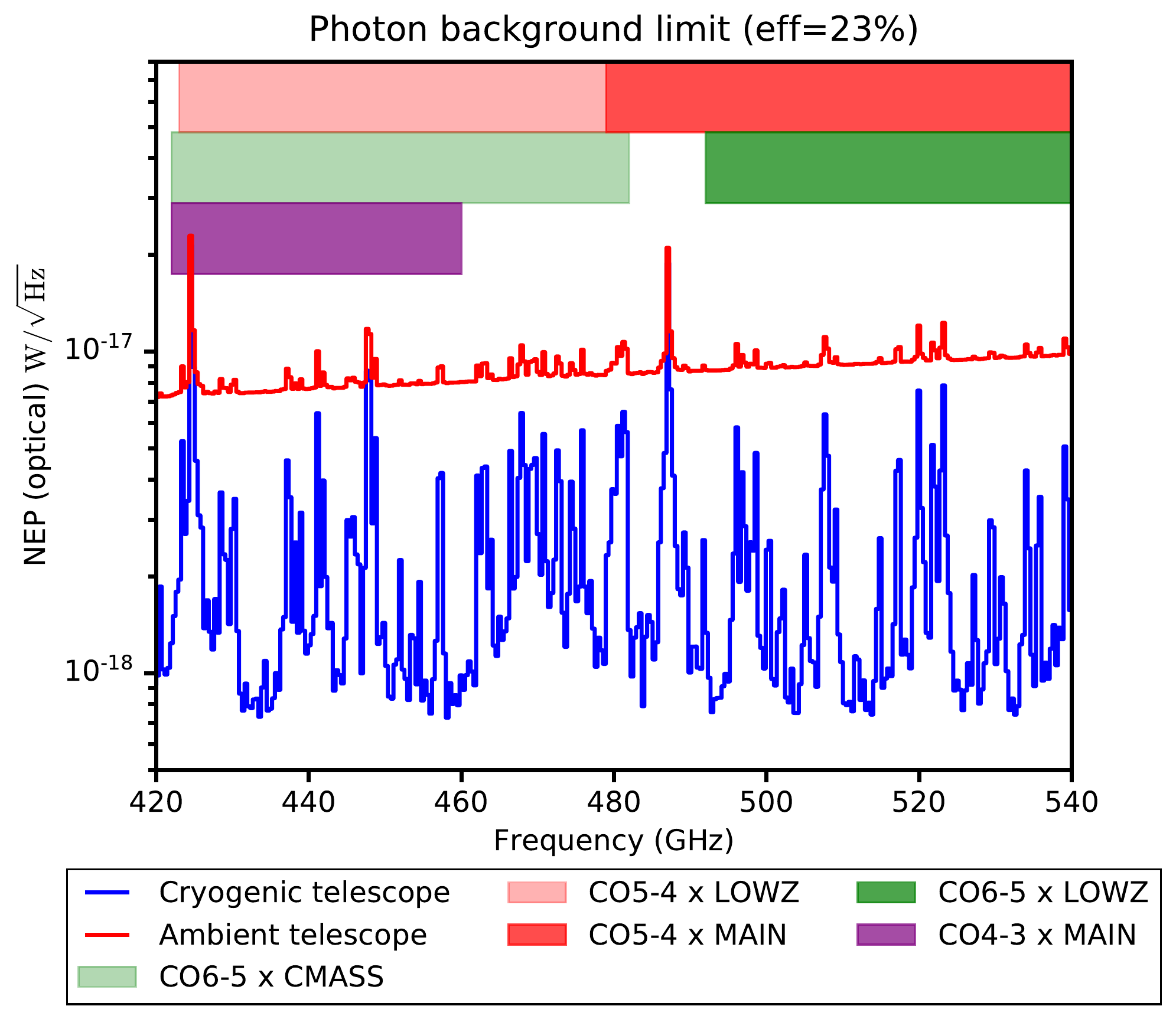}
\end{tabular}
  \end{center}
  \caption[Optics]{Photon loading (left) and background-limited optical NEP (right), both measured with respect to power per spectrometer channel defined at the input to the main lobe of the spectrometer lenslet beam and assuming a telescope optical efficiency up to this point of $74\%$. The background limit includes optically-excited quasiparticles and assumes $23\%$ spectrometer efficiency. The column depth and pressure broadening (${\approx}10$~MHz/Torr) of the atmosphere drop dramatically from the ground to the altitude of the balloon at float. In broad photometric bands, the upper atmospheric emission is similar to one bounce from a mirror or transmission through a window at ambient temperature. The diminished pressure broadening also opens windows that are much darker than the broadband average, and which become accessible to a spectrometer with resolving power $R > 100$. Observations in these windows also benefit from cryogenic optics; over \exclaimband\ with $R=512$, $50\%$ of the channels have radiation background 25 times lower than an ambient-temperature optic.}
  \label{fig:photon_background}
\end{figure*}

EXCLAIM's scientific goals can be summarized as follows: 1) to make a definitive detection of redshifted [CII]~\cite{2019MNRAS.489L..53Y, 2019MNRAS.488.3014P} ($1900$~GHz rest frame) in correlation with quasars at redshifts $2.5 < z < 3.5$; 2) to detect two adjacent ladder lines of CO in each of the BOSS samples (MAIN, LOWZ, CMASS); and 3) to constrain both CO $J=4-3$ and [CI] ($492$~GHz) in the Milky Way. Current models of the diffuse, redshifted CO and [CII] intensity vary by nearly two orders of magnitude~\cite{2013ApJ...768...15P, 2011ApJ...741...70L, 2008A&A...489..489R, 2010JCAP...11..016V, 2016MNRAS.461...93P, 2020arXiv200911933Y}. EXCLAIM's IM measurements of two $J$ lines at the same redshift will permit new diagnostics of the ISM environment. Additionally, EXCLAIM's measurements of the integrated emission will complement interferometric observations by the Atacama Large Millimeter/submillimeter Array (ALMA)~\cite{2019ApJ...882..138D}. In the Milky Way, [CI] emission can track molecular gas in regions where CO is photo-dissociated\cite{2015ApJ...811...13B, 2010ApJ...716.1191W}. Thus, EXCLAIM will provide insight into the relation between CO and molecular gas. 

EXCLAIM uses an all-cryogenic ($<5$~K) telescope that provides the high sensitivity required to achieve these goals in a one-day conventional balloon flight from North America (i.e., Texas or New Mexico). These North American launch sites provide excellent access to the BOSS regions and offer easy logistics and reuse. Figure~\ref{fig:photon_background} shows the low photon loading expected at the input to the spectrometer lenslet main beam. The combination of a cold telescope with a moderate-resolution spectrometer, observing from a stratospheric balloon, will allow EXCLAIM to observe with sufficient sensitivity in the dark regions between atmospheric lines\cite{2019zndo...3406483P}, which are broadened into a continuum at lower altitudes. 

The EXCLAIM program began in April $2019$ and is approaching the end of its preliminary design and technology completion stage, with several systems already in their final design phase. An engineering flight to test all the systems with one spectrometer is planned in $2022$. Following refurbishment, science flights with six spectrometers will start as early as $2023$. This manuscript describes each major system's design and concludes with detection forecasts.

\section{Cryogenic telescope}
\begin{figure*}[t!]
\begin{center}
    \includegraphics[width=.85\textwidth]{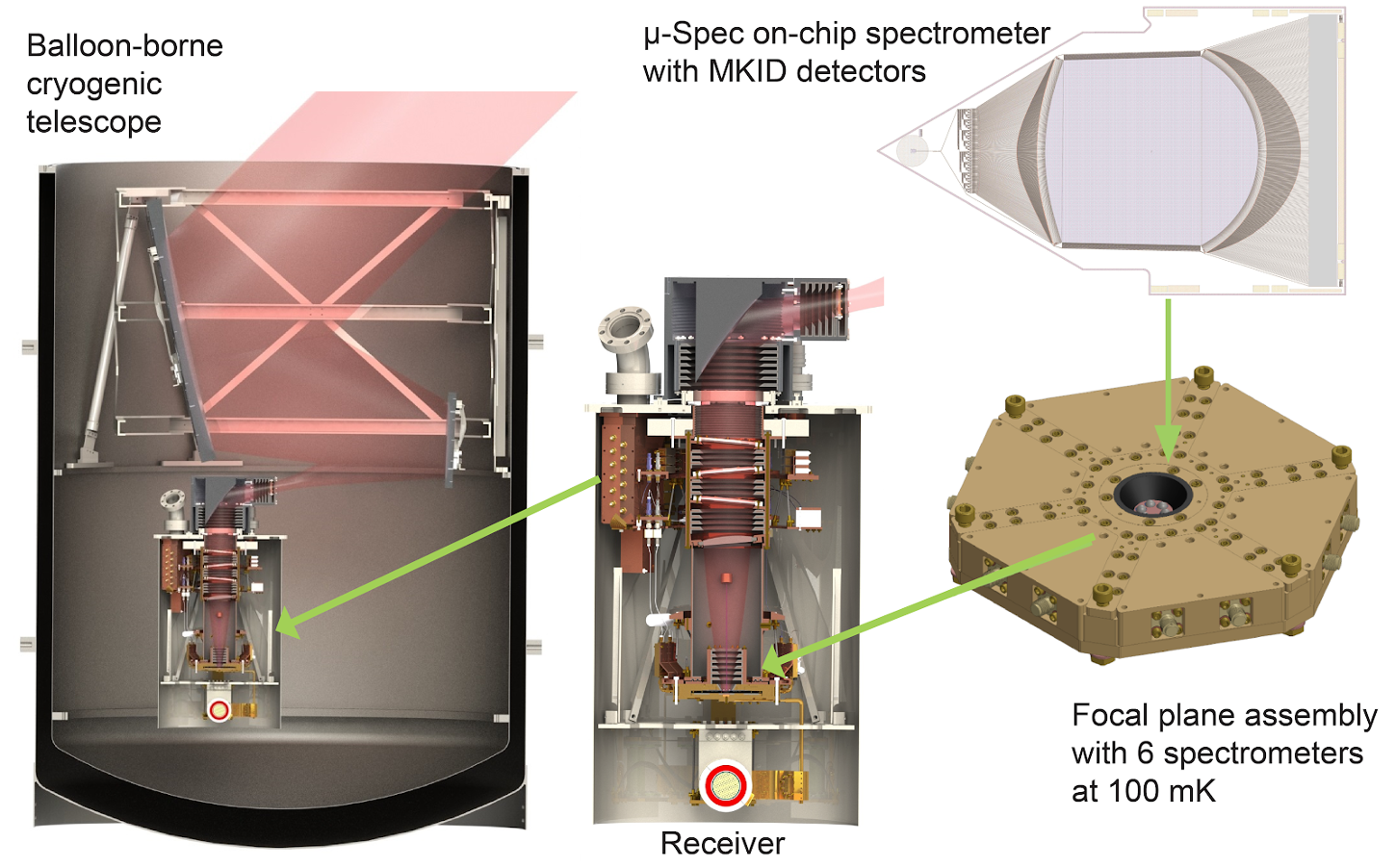}
  \end{center}
  \caption[Optics]{EXCLAIM employs a cryogenic telescope to reach high sensitivity in spectral windows of low emission in the upper atmosphere. The telescope and receiver are located in a bucket dewar that launches with ${\approx}2500$~liters of liquid-helium to provide $24$ hours of $1.7$~K operations at float altitude. The receiver houses the focal plane assembly with six spectrometers operating at $\approx 100$~mK.
  }
  \label{fig:exclaim_overview}
\end{figure*}

As shown in Fig.~\ref{fig:exclaim_overview}, infrared radiation enters the telescope at an elevation angle of $45^\circ$ and is reflected off of a $90$~cm ($76$~cm projected) monolithic aluminum parabolic primary mirror~\cite{2016SPIE.9914E..1KH, 2016SPIE.9914E..1JG} to a flat mirror, and then onto a secondary mirror which produces a collimated beam that couples to the receiver. The projected aperture size provides an angular resolution of $4.2'$~full width at half maximum (FWHM), permitting a survey that covers spatial scales from the linear regime, $k \lesssim 0.1 h/{\rm Mpc}$, up to scales where shot noise dominates, $k \gtrsim 5 h/{\rm Mpc}$~\cite{2019arXiv190710067B}. For more information on the optics design, see Ref.~\citenum{Essinger-Hileman2020} in these proceedings. The cryogenic receiver houses the focal plane assembly with six spectrometers operating at $\approx 100$~mK.

The EXCLAIM telescope and receiver are housed in a liquid-helium Dewar, with an inner diameter of $152$~cm and identical in design to the Absolute Radiometer for Cosmology, Astrophysics, and Diffuse Emission II (ARCADE II)~\cite{2011ApJ...730..138S} and the Primordial Inflation Polarization ExploreR (PIPER)~\cite{2016SPIE.9914E..1JG} instruments. An initial fill of ${\approx} 2500$~liters provides $14$~hours of $1.7$~K science operation at a float altitude greater than $27$ km. A swiveling lid insulates the instrument on the ground and in ascent. At float altitude, positive pressure from boil-off gas acts to keep the cryogenic optics dry and clean, eliminating the need for windows at ambient temperature. ARCADE~II and PIPER have similar open aperture areas and have demonstrated positive pressure. PIPER engineering flights~\cite{2018SPIE10708E..06P} showed that superfluid fountain-effect pumps could maintain large telescope optical elements at $1.7$~K.


\section{Receiver}
The receiver houses the spectrometers, amplifiers and sub-Kelvin cooling system (Fig.~\ref{fig:exclaim_overview}), and will remain superfluid-tight. Light enters the receiver through a anti-reflection (AR)-coated silicon window and is brought to a focus with an AR-coated silicon lens. Hyper-hemispherical lenslets~\cite{1993ITMTT..41.1738F, 2000stt..conf..407J} placed over the on-chip spectrometer slot antennas couple the light to the spectrometer chip. The input optical path is collimated, permitting a long baffle region and two layers of magnetic shielding with high aspect ratio. Cryogenic housekeeping harnesses and the spectrometer readout are routed to ambient-temperature electronics via thin-wall stainless-steel bellows. 


A single-shot Adiabatic Demagnetization Refrigerator (ADR)~\cite{10.1117/1.JATIS.4.2.021403, 2019RScI...90i5104S} provides a $100$~mK base temperature and ${\approx}1$~\textmu W of cooling power for the spectrometer focal plane. Niobium-titanium (NbTi) coaxial cables~\cite{2017ITAS...2800105W} and carbon fiber suspensions isolate the $100$~mK stage from a $900$~mK intermediate stage, which is cooled by a $^4{\rm He}$ adsorption cooler. The high-current ADR magnet leads are vapor-cooled and enter the receiver through a superfluid-tight feedthrough.

\section{Integrated spectrometer}
\label{sec:spectrometer}

\begin{figure*}[!t]
    \begin{center}
        \includegraphics[width=0.85\textwidth]{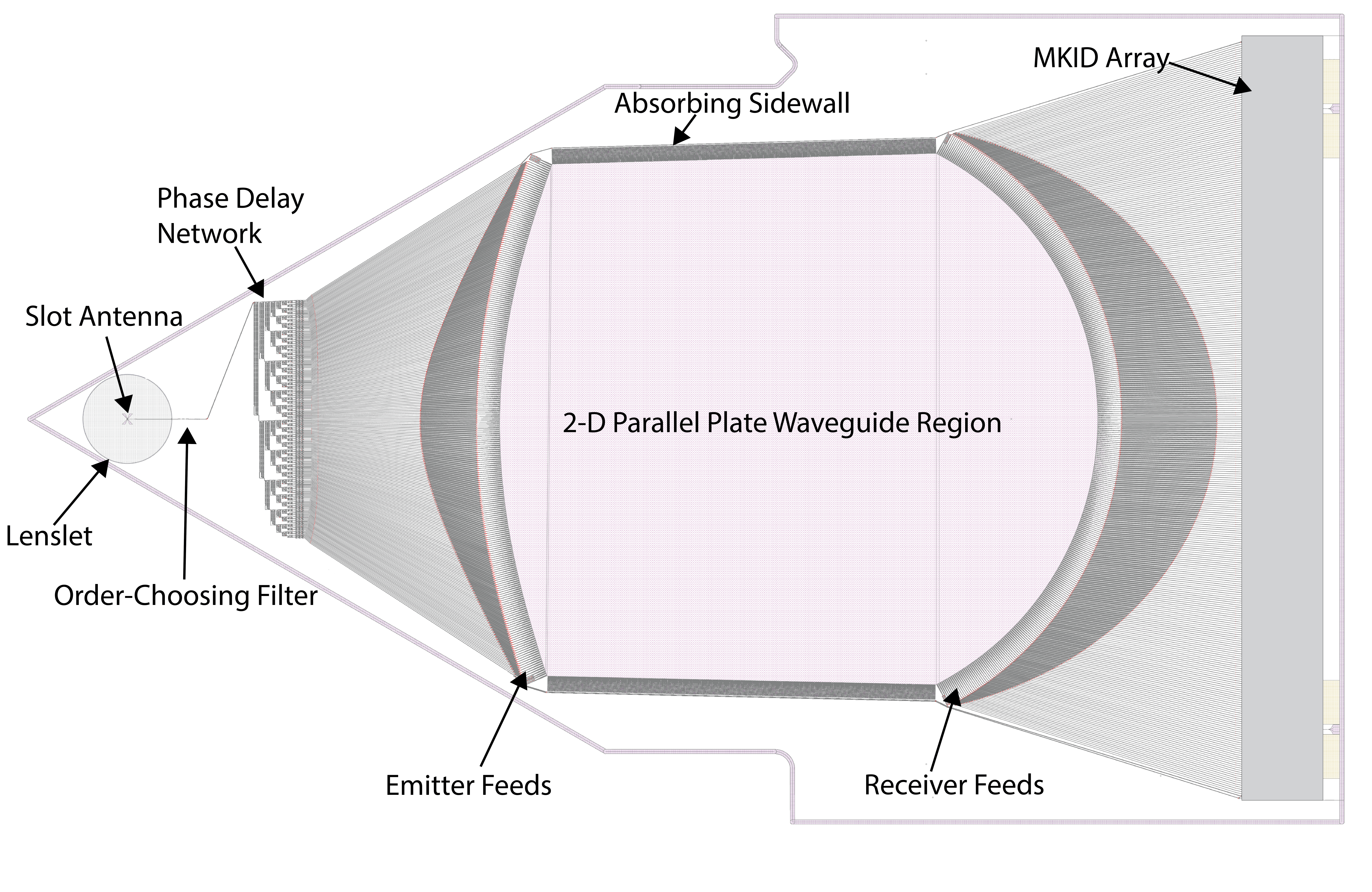}
       \caption[\mspec\ device]{The EXCLAIM spectrometer design. The light enters the spectrometer through a slot antenna and propagates through a microstrip transmission-line delay network before being launched in a 2D parallel plate waveguide region by antenna feed structures. The light diffracts and the spectrum is Nyquist-sampled by another array of antenna feed structures, which are connected to a bank of MKIDs. An order-choosing filter placed past the slot antenna selects the grating order in use.}
       \label{fig:r512photo}
   \end{center}
\end{figure*}

The focal plane of the EXCLAIM instrument employs six \mspec\ \cite{2014ApOpt..53.1094C, CATALDO201554, barrentine2016design} spectrometers. \mspec\ is a diffraction-grating analog spectrometer, which integrates all of its components, including the detectors, onto a compact silicon (Si) chip (see Fig.~\ref{fig:r512photo}).

A slot-antenna with an on-chip Si lenslet couples the light from the receiver optics to planar niobium (Nb) transmission lines and a diffraction grating is synthesized in a microstrip line delay network. This network introduces a linear phase gradient and then launches the signals into a Nb 2D parallel-plate waveguide region via an array of emitting and receiving feeds that are arranged in a Rowland configuration~\cite{rowland1883,rotman1963wide}. These receivers Nyquist-sample the spectrometer's spectral function, which is close to a ${\rm sinc}^{2}$ function, the Fourier transform of a uniformly illuminated synthetic grating. Each received signal is then transmitted to individual microwave kinetic inductance detectors (MKIDs).

The thin Si dielectric layer and protective Nb ground plane provide high immunity to stray light and cross-talk. In addition, a thin-film titanium (Ti) coating is deposited on the back of the spectrometer chip to terminate stray light and thermal blocking filters~\cite{4751534, 2014RScI...85c4702W} can be placed on the microwave input and output lines. Each EXCLAIM spectrometer employs an array of $355$ aluminum (Al) microstrip transmission line MKIDs, operating at near background-limited sensitivity and read out on a single microwave readout line.

The fabrication of the spectrometer circuitry on the single-crystal Si dielectric uses a wafer-scale bonding technique~\cite{patel2013fabrication}. A Nb liftoff patterning process is used, which provides precision control of line width~\cite{patel2013fabrication} to ensure the necessary phase control in the spectrometer transmission lines and avoids damaging the thin Si dielectric substrate. Single-crystal Si has low-loss~\cite{Wollack:20,cataldo2015analysis,kongpop2016cryogenic}, which enables near-unity efficiency and resolutions~\cite{cataldo2018design, barrentine2016design} up to $R=1500$ in principle. In practice, the EXCLAIM spectrometer efficiency is expected to be limited not by transmission-line loss, but by the individual efficiencies of the sub-millimeter component designs, with a total expected spectrometer efficiency around $23\%$. For more details of the EXCLAIM spectrometer design and fabrication, see Ref.~\citenum{Mirzaei2020} in these proceedings.

\section{Instrument electronics and flight software}
The EXCLAIM detector readout electronics baseline design includes a Xilinx ZCU111 Radio Frequency System-on-Chip (RFSoC). An ongoing trade study shows clear advantages in terms of performance, mass and power over a Reconfigurable Open Architecture Computing Hardware 2 (ROACH-2) system~\cite{2016JAI.....541001H}, which was employed by previous balloon missions such as The Next Generation Balloon-borne Large Aperture Submillimeter Telescope (BLAST-TNG)~\cite{2018SPIE10708E..0LL} and the Far Infrared Observatory Mounted on a Pointed Balloon (OLIMPO)~\cite{2019JCAP...01..039P}. Intermediate frequency hardware mixes the digital-to-analog converter (DAC) output from either an RFSoC or ROACH-2 to the ${\approx}3.5$~GHz MKID resonance frequency range and mixes the spectrometer output down for input to the analog-to-digital converter (ADC). The readout provides $512$~MHz radio frequency (RF) bandwidth and reads the spectrometer array at $488$~Hz. 

The electronics for the ADR, cryogenic and gondola housekeeping, attitude determination and control, power control, and telemetry interface are based on the PIPER design~\cite{2018SPIE10708E..06P}, although EXCLAIM's unique requirements dictate several variations. The flight software and low-power flight computers also follow the PIPER design approach, demonstrated in the PIPER $2017$ engineering flight~\cite{2018SPIE10708E..06P}.

\section{Thermal control}
\begin{figure}[!t]
    \begin{center}
        \includegraphics[width=0.48\textwidth]{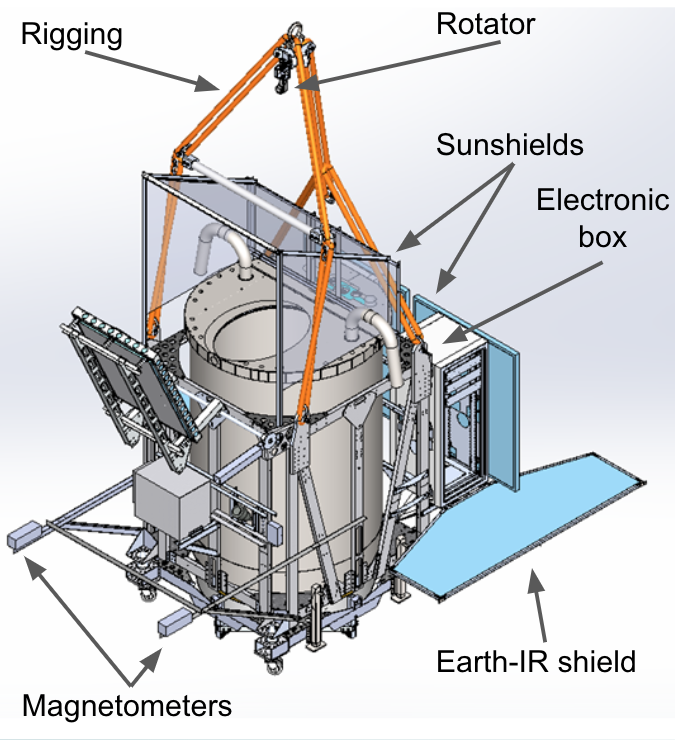}
        \caption[Gondola]{The complete payload. A rotator pivot connects EXCLAIM to the flight train of the balloon through a rigging that holds the entire payload. The different shields enable passive thermal control. The magnetometers serve as attitude determination sensors along with a star camera, gyroscopes, sun sensors and clinometers (not shown).}
        \label{fig:gondola}
    \end{center}
\end{figure}

Except for the receiver, the thermal control of the instrument is done passively, although localized heaters can be used as needed. The flight electronics (Fig.~\ref{fig:gondola}) are stored in a $33$U rack, while the detector readout electronics are stored in a $14$U rack.\footnote{A rack unit, or ``U", is a unit of measure defined as $1.75$ inches ($44.45$~mm). It is used as a measurement of the overall height of $19$- and $23$-inch-wide rack frames.} Both racks are $19$~inches wide.
In the current design, the batteries are located underneath the $14$U rack to ensure the payload is balanced.

In order to maintain the electronics temperature between $-40^\circ$C and $+20^\circ$C, several radiators function to dissipate excess heat. Multilayer insulation (MLI) radiation shields are placed around the payload (Fig.~\ref{fig:gondola}) to reduce the amount of incident solar and Earth infrared radiation. In addition, a sunshield is added above the lid to reduce solar radiation on the telescope aperture during daytime operations.
Foam panels placed around the walls of the electronics box and heaters keep the electronics warm during ascent, when temperatures can drop below $-50^\circ$C.

\section{Observing strategy}
\label{sec:ADCS}
EXCLAIM will scan in azimuth at constant elevation and map a strip across a constant range of declination. Unlike the auto-power~\cite{1995PhRvD..52.4307K}, a cross-power detection does not suffer a penalty toward large survey areas. Thus, EXCLAIM seeks to maximize the survey area to access linear modes while also sampling the beam in the scan and sky drift directions. For a flight from Ft. Sumner, New Mexico, the provisional extragalactic survey region encompasses Sloan Stripe $82$.
Primary science observations will occur when the sun is down, but additional galactic regions may be achievable in an anti-solar scan pattern during the day.

EXCLAIM has modest pointing requirements to conduct a large-area survey within the BOSS regions. A reaction wheel scans the payload about a central azimuth to enable the primary mapping strategy. A rotator pivot to the flight train provides torque to desaturate the reaction wheel by dumping momentum to the balloon. The reaction wheel and rotator are sized to provide control torque at least equal to worst-case disturbances. A magnetometer and a star camera determine an instantaneous reference to the scan center azimuth. Sun sensors are used primarily for sun avoidance during the day time, but can also be used as a cross-check on the magnetometer for absolute pointing determined in-line. Post-flight pointing reconstruction uses data from an array of gyroscopes, accelerometers, tilt sensors and the magnetometer to tie together star camera determinations, which will provide $\approx 3"$ pointing. 

\section{Anticipated sensitivity}
\label{sec:forecast}

The sensitivity for intensity mapping is estimated using both a numerically simple mode counting argument for the three-dimensional power spectrum~\cite{2011ApJ...741...70L, 2014ApJ...786..111P, 2016ApJ...817..169L} and a simulated analysis of angular cross-correlations between redshift slices~\cite{2019MNRAS.485..326L}. Both estimates agree and include the effects of angular and spectral resolution and survey volume. The three-dimensional power spectrum approach requires homogeneous noise in the frequency direction and an inverse-noise weighted effective noise is used for the volume, which is validated by simulations. Published survey depths~\cite{2016MNRAS.455.1553R, 2015MNRAS.453.2779E} give shot noise in the sample.

\begin{figure*}
    \begin{center}
    \begin{tabular}{ll}
        \includegraphics[width=0.47\textwidth]{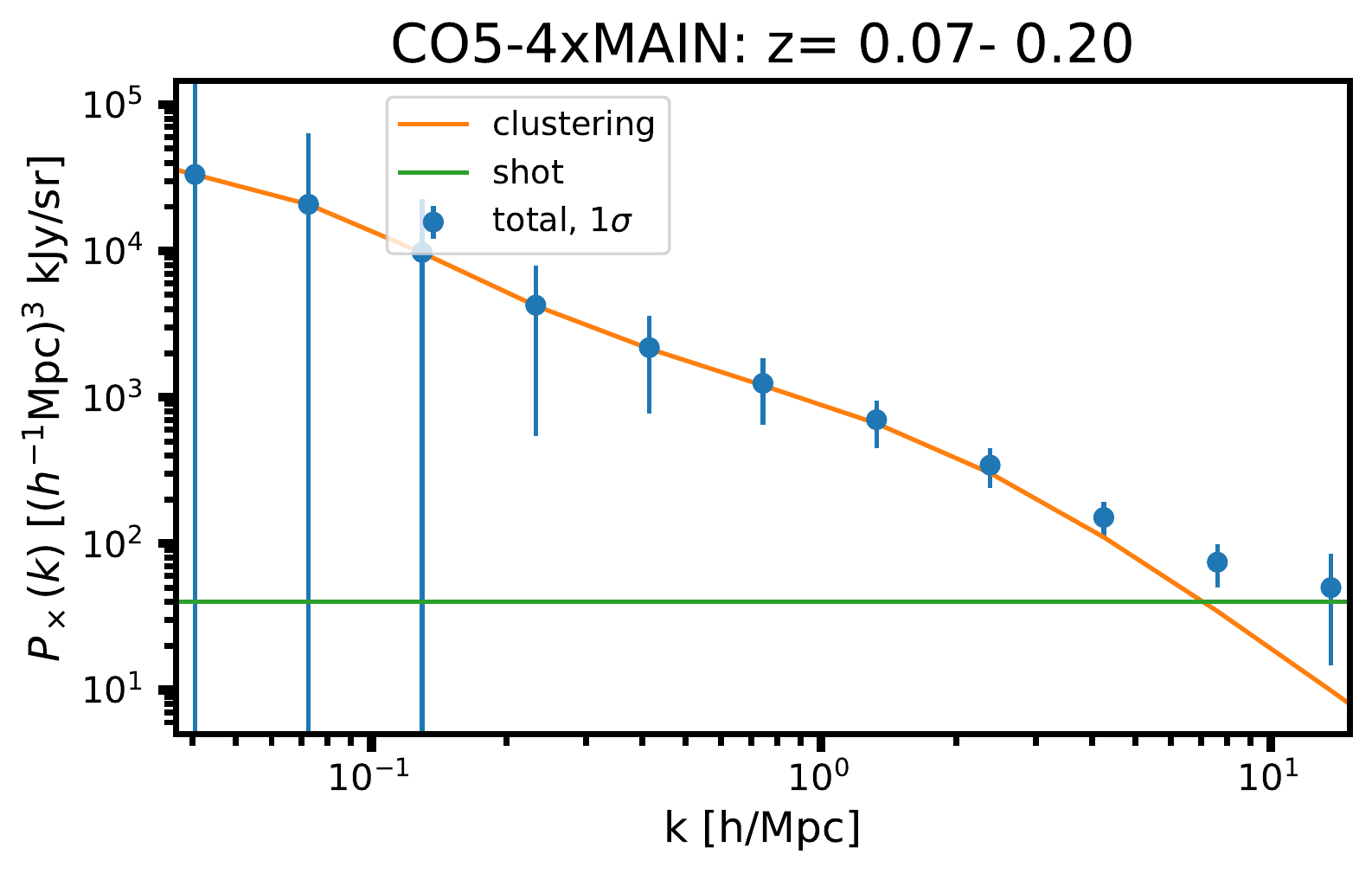} &
        \includegraphics[width=0.47\textwidth]{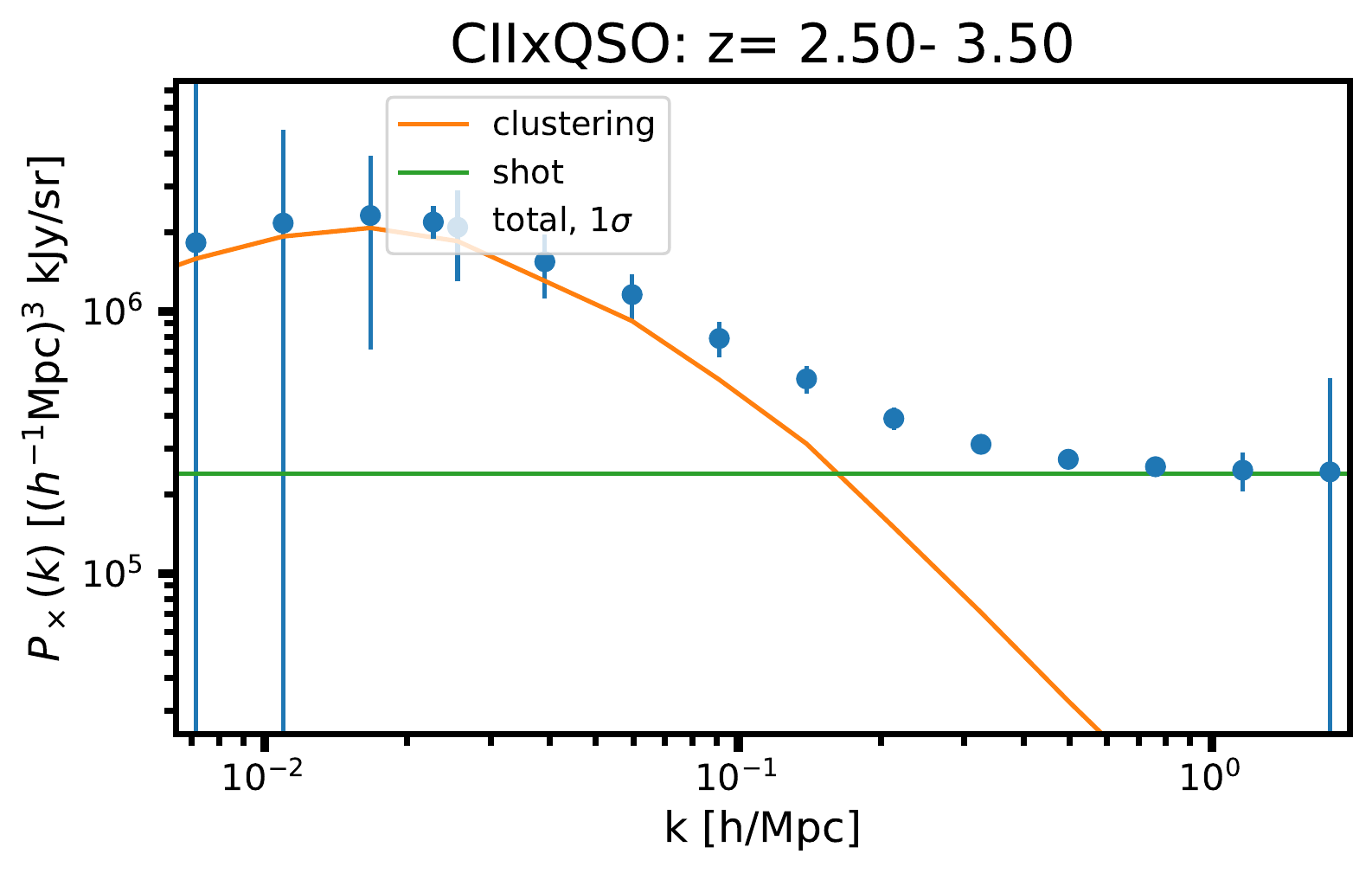}
    \end{tabular}
  \end{center}
  \caption[Forecast]{Forecast for cross-correlations of CO $J=5-4$ $\times$ SDSS MAIN at $0.07<z<0.2$ (left) and [CII] $\times$ BOSS QSO at $2.5<z<3.5$ (right) assuming signal levels from Refs.~\citenum{2013ApJ...768...15P, 2019MNRAS.489L..53Y}, respectively. The low redshift correlations trace nonlinear scales and are expected to have limited correlated shot noise. The high-redshift [CII] probes linear scales and is expected to have a higher level of correlated shot noise.}
  \label{fig:forecast}
\end{figure*}

The specification of the detector performance for forecasts must take several factors into account: 1) MKID noise contributions are loading-dependent, 2) the power absorbed by the MKID depends on the efficiency of the complete spectrometer system, and 3) MKIDs generically have $1/f$ noise contributions from two-level systems~\cite{2008ApPhL..92u2504G}. To accommodate these factors, in these forecasts the sensitivity of the spectrometer is specified under expected optical loading as a multiplier of the photon background-limit (including optically-excited quasiparticle fluctuations), referring to the power incident at the spectrometer lenslet, and weighted over acoustic frequencies of the science signal ($5$--$25$~Hz, unless additional modulation is employed). The spectrometer NEP for these baseline forecasts is conservatively taken to be a factor of three over the background limit, though current MKID design sensitivity is nearer to the background limit. Conversely, poorer NEP performance will still accomplish the EXCLAIM mission threshold detection goals. Future publications will describe the loading-dependent noise model for the MKID design, forecast science, and margins. As shown in Fig.~\ref{fig:forecast}, the expected $2\sigma$ sensitivity to the surface brightness-bias product for $0<z<0.2$ (SDSS MAIN) for CO $J=4-3$, $J=5-4$, $0.2<z<0.4$ for $J=5-4$, $J=6-5$ (BOSS LOWZ), $0.4<z<0.7$ for $J=6-5$ (CMASS), and $2.5<z<3.5$ for [CII] (QSO) are $\{ 0.15, 0.28, 0.30, 0.37, 0.45, 13 \}$~kJy/sr, respectively.


\section{Conclusions}
In this manuscript we have described the EXCLAIM mission and how this balloon-borne cryogenic telescope will provide insight into star formation history over cosmic time. EXCLAIM will use a line intensity mapping approach to perform a complete census of the emitting gas, creating a map of all emission in a given line, free from selection bias. The instrument design, enabled by a cryogenic telescope and integrated MKID spectrometers will provide near background limited sensitivity to detect CO and CII emission lines from redshifts of $0<z<3.5$. EXCLAIM is slated to launch for an engineering flight in late summer or fall of 2022, followed by science flights starting 2023.

\section*{Acknowledgments}       
Funding provided by the NASA Astrophysics Research and Analysis (APRA) program is gratefully acknowledged, as well as funding from the Space Grant in support of several EXCLAIM interns.

\bibliography{references}
\bibliographystyle{spiebib} 
\label{bib}

\end{document}